\documentclass{iaus}
\usepackage{graphicx}

\title[VMC Planetary Nebulae] 
{Planetary Nebulae in the\\ VISTA Magellanic Cloud (VMC) Survey}

\author[B. Miszalski et al.] 
{B. Miszalski$^{1,2}$, R. Napiwotzki$^{3}$, M.-R.L. Cioni$^{3,4}$, M.A.T. Groenewegen$^{5}$, J.M. Oliveira$^{6}$, A. Udalski$^{7}$ \& J. Nie$^{8,9}$
}

\affiliation{
$^1$South African Astronomical Observatory\ 
$^2$Southern African Large Telescope Foundation
\\email: {\tt brent@saao.ac.za}\\[\affilskip]
$^3$University of Hertfordshire\ 
$^4$University Observatory Munich\ 
$^5$Royal Observatory of Belgium\\[\affilskip] 
$^6$Keele University\
$^7$Warsaw University Observatory\ 
$^8$Australian National University\\[\affilskip]
$^9$Beijing Normal University
}

\pubyear{2011}
\volume{283}  
\pagerange{1-2}
\jname{Planetary Nebulae an Eye to the Future}
\editors{A.C. Editor, B.D. Editor \& C.E. Editor, eds.}
\begin{document}

\maketitle

\begin{abstract}
   The multi-epoch $YJK_s$ sub-arcsecond photometry of the VMC survey provides a long anticipated deep near-infrared (NIR) window into further understanding the stellar populations of the Magellanic Clouds. The first year of observations consisted of six tiles covering $\sim$9\% of the Large Magellanic Cloud (LMC) survey region and contains 102 objects previously classified as planetary nebulae (PNe). A large proportion of the sample were found to be contaminated by non-PNe. These initial results underline the importance of establishing a clean catalogue of LMC PNe before they are applied in areas such as the planetary nebula luminosity function (PNLF) and searches for binary central stars. As the VMC survey progresses it will play a fundamental role in cleaning extant PN catalogues and a complementary role in the discovery of new PNe. 

\keywords{planetary nebulae: general, Magellanic Clouds}
\end{abstract}

\firstsection 
              
\section{Introduction}
Magellanic Cloud PNe (MCPNe) are the nearest large population of extragalactic PNe where the low reddening and favourable viewing angle allows for population-wide studies not possible in the Milky Way. The most important of these is the extragalactic standard candle [O~III] PNLF in which MCPNe serve as the benchmark population (Ciardullo et al. 2010). However, until a detailed multi-wavelength analysis of MCPNe to remove contaminating non-PNe is performed, their acclaimed benchmark status could be considered uncertain. The deep $YJK_s$ photometry of the VMC survey will allow for a thorough appraisal of the majority of MCPNe for the first time.

\firstsection 
\section{A multi-wavelength study of 102 LMC PNe and non-PNe}
\begin{figure}
   \centering
      \includegraphics[scale=0.7]{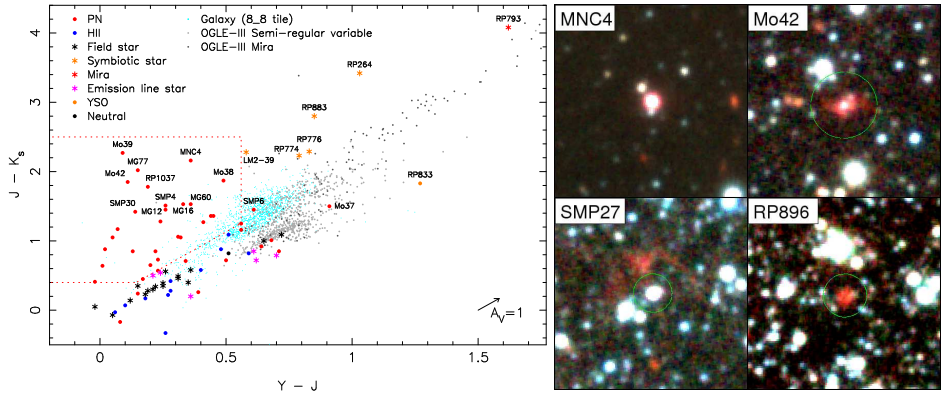}
      \caption{\emph{(left)} The VMC `ant diagram' showing the position of PNe and various non-PNe in the sample analysed by Miszalski et al. (2011b). \emph{(right)} Colour-composites of four bona-fide PNe made from stacked $K_s$ (red), $J$ (green) and $Y$ (blue) VMC images.}
   \label{fig:fig1}
\end{figure}

Miszalski et al. (2011b) performed a multi-wavelength analysis of 102 objects using VMC data (Cioni et al. 2011) and a host of mid-infrared and optical observations. A total 46/67 or 69\%\footnote{This includes RP227 mistakenly omitted from Table 6 of Miszalski et al. (2011b).} of objects in our sample from Reid \& Parker (2006b) were reclassified as non-PNe (HII regions, field stars, emission line stars, symbiotic stars and a young stellar object). These results are numerically dominated by the complex 30 Doradus region and some reclassifications may be explained by the relatively low survey resolution of Reid \& Parker (2006a). Miszalski et al. (2011b) developed a range of diagnostic diagrams (e.g. Fig. \ref{fig:fig1}) to guide future analysis of larger samples of MCPNe with the growing VMC dataset (Miszalski et al. in preparation). 

The inclusion of OGLE-III $I$-band (Udalski et al. 2008a,b) and VMC $K_s$ lightcurves were a powerful means to firmly identify contaminating variable stars (e.g. the Mira RP793 with $\Delta K_s=\sim$0.4 mag and 503.4 day period from OGLE-III). None of the variables identified had the multi-wavelength properties of typical PNe, which underscores the importance of a clean PN population before searching for binary central stars. The large fraction of giant companions claimed by Shaw et al. (2009) is instead explained by the numerous field stars and emission line stars found by Miszalski et al. (2011b). 

\firstsection 
\section{Towards an improved census of LMC PNe}
The Reid \& Parker (2010) [O~III] PNLF of LMC PNe is a substantial advance over previous work, however further effort is required to improve this important PNLF. Firstly, only a small fraction of the sample has been cleaned of non-PNe and secondly, there remain many undiscovered PNe to be added. Miszalski et al. (2011a) found 2--3 new PNe in a $63\times63$ arcmin$^2$ region not listed by Reid \& Parker (2006b) and concluded that perhaps 50--75 new PNe remain to be found in the inner $5\times5$ deg$^2$ of the LMC. Reid \& Parker (these proceedings) report on new PNe found outside this zone using optical selection criteria. Optical searches may not find all PNe as demonstrated by MNC4 (Miszalski et al. 2011a) whose [WC9] or later central star is too cool to ionise [O~III]. With more VMC data we will build NIR PNLFs to complement the [O~III] PNLF.

\end{document}